\documentclass[11pt]{article}
\date{}

\usepackage[left=25mm, right=25mm, top=25mm, bottom=25mm, includehead=false, includefoot=false]{geometry}

\usepackage{graphicx}
\usepackage{url}
\usepackage[round,semicolon]{natbib}  
\usepackage{authblk} 

\usepackage[table]{xcolor}
\usepackage[parfill]{parskip} 
\usepackage{amsmath}
\usepackage{sectsty}
\allsectionsfont{\sffamily}

\usepackage[T1]{fontenc}

\usepackage[pdftex]{hyperref} 
\hypersetup{pdfborder={0 0 0} }

\title{\sffamily\fontsize{16}{0}\textbf{Understanding population fluctuations through volunteered geographic information and novel indicators: The experience of Rakiura, Stewart Island, New Zealand}}

\author[1]{Mathew Darling\thanks{}}
\author[2]{Benjamin Adams}
\author[3]{Caroline Orchiston}
\author[1]{Thomas Wilson}
\author[4]{Brendon Bradley}

\affil[1]{Department of Geological Sciences, University of Canterbury}
\affil[2]{Department of Geography, University of Canterbury}
\affil[3]{Centre for Sustainability, University of Otago}
\affil[4]{Civil and Natural Resource Engineering, University of Canterbury}
\affil[*]{\texttt{Email: mathew.darling@pg.canterbury.ac.nz}}

\begin{document}

\maketitle


\begin{abstract}
\noindent
\setlength{\parindent}{0pt}

In an era of heterogeneous data, novel methods and volunteered geographic information provide opportunities to understand how people interact with a place. However, it is not enough to simply have such heterogeneous data, instead an understanding of its usability and reliability needs to be undertaken. Here, we draw upon the case study of Rakiura, Stewart Island where manifested passenger numbers across the Foveaux Strait are known. We have built a population model to ground truth such novel indicators. In our preliminary study, we find that a number of indicators offer the opportunity to understand fluctuations in populations. Some indicators (such as wastewater volumes) can suggest relative changes in populations in a raw form. While other indicators (such as TripAdvisor reviews or Instagram posts) require further data enrichment to get insights into population fluctuations. This research forms part of a larger research project looking to test and apply such novel indicators to inform disaster risk assessments.

$ $ 
\end{abstract}


\section{Introduction}
Globally inbound tourism markets are volatile; significant changes can be sensitive to political unrest, natural disasters or other crises, leading to reduced visitor arrivals and slow recovery. New Zealand has a relatively small dispersed population with proportionally high tourist arrivals, and understanding volatility is critical for infrastructure and natural hazard management \citep{orchiston2012seismi}. The vulnerability of  tourism destinations to natural disasters is high, as it ``relies so heavily on perceptions of safety, functioning infrastructure and visitor mobility'' (p. 59). Moreover, due to the possibility of a negative feedback cycle developing (i.e., following a natural disaster a tourist will not visit a specific country or region \citep{orchiston2013insurance}) it is important that dynamic nature of transient populations is better understood. 

In this study we consider the use of volunteered geographic information (VGI) and novel population movement indicators as predictors of tourist travel patterns to support disaster risk reduction. The potential for VGI and crowdsourced datasets as sources of information to support disaster response is well-studied, yet its quality and usefulness remain debatable for many scenarios \citep{goodchild2010crowdsourcing,goodchild2012assuring,senaratne2017review}. In prior work, a common motivation for using VGI and novel data sources for population mapping has been to understand risk exposure to infectious disease in places without accurate population figures \citep{deville2014dynamic,tatem2012mapping}. In our case we consider a different situation where the population numbers are influenced by a large number of transient tourists and thus are highly dynamic.

A case study area, of Rakiura, Stewart Island, New Zealand's southern-most inhabited island has been selected. Here, confidence in the actual population movement can be derived from passenger manifests on the ferry and air services, visitor levy data collected by the local council, and international cruise ship positioning data (automatic identification system). We compare this population model to VGI and novel indicators on two temporal scales. These methods are tested to see whether such indicators can be adopted as a proxy for population fluxes, and in turn be used to better inform disaster risk assessments.

\begin{figure}[htbp] \begin{center} 
\resizebox{1.0\textwidth}{!}{ 
	\includegraphics{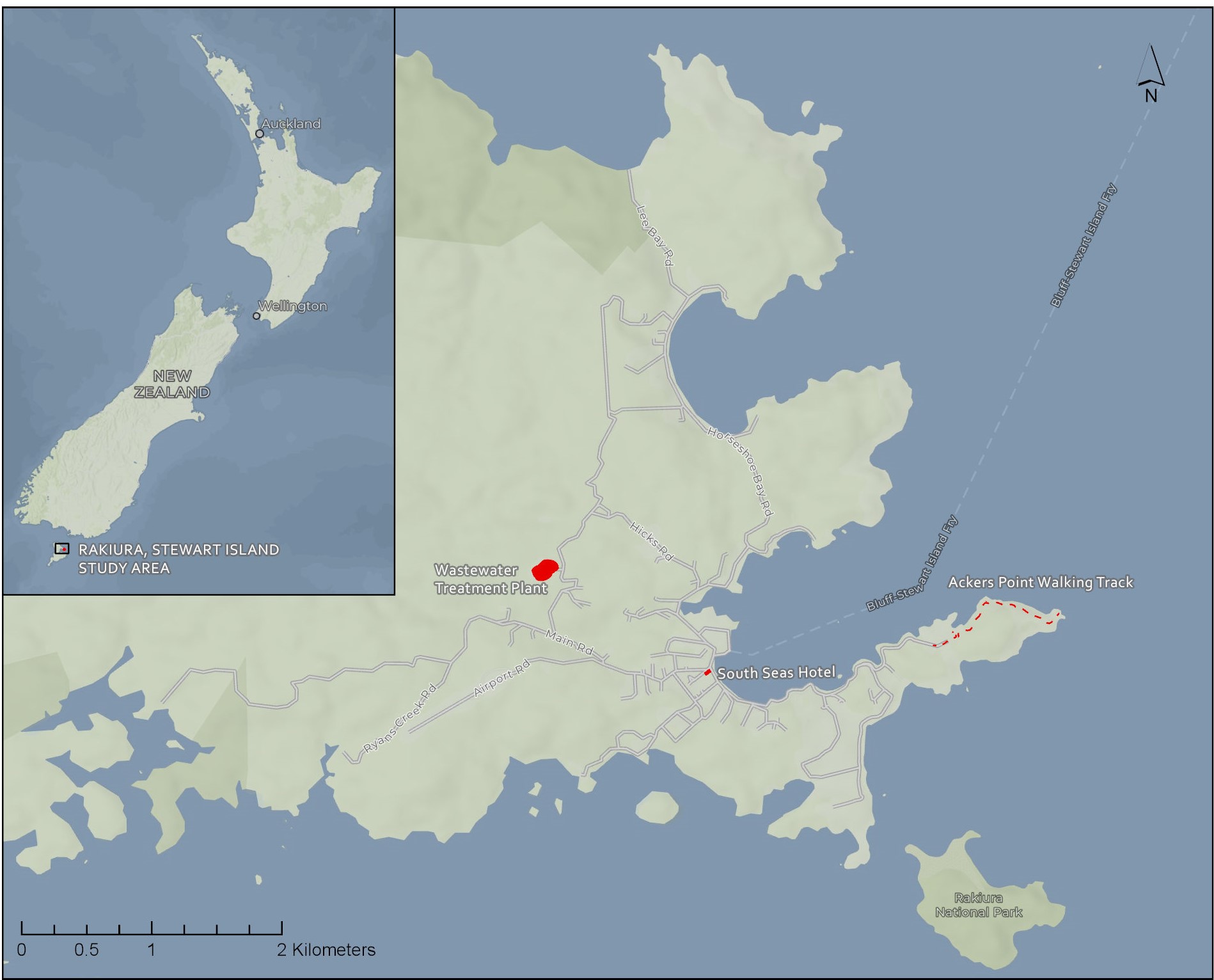}
} \caption{Case Study Area of Stewart Island Rakiura, and location of novel/VGI indicators} \label{first_figure} \end{center} \end{figure}

In summary, the aims of this study are two-fold. Firstly, it is to understand how well VGI and novel indicators represent fluctuations in population. Secondly, it considers opportunities to undertaking data enrichment methods to build confidence in the assessed indicator.

\section{Methodology}

\subsection{Modelling the population of Rakiura, Stewart Island}
Similar to wider New Zealand, Stewart Island (or \textit{Rakiura} in M\=aori), has experienced significant growth in tourism in recent years. International and domestic tourists alike enjoy visiting the island's wilderness and natural areas. As such, the population (378, as of 2013) can increase three-fold with a visiting international cruise ship, or during a busy public holiday period. Fluctuations in the island's populations are largely driven by seasonality, with the busiest periods being between September and March, coinciding with more favourable weather, a number of public holidays, and the cruise `season'.   

\subsubsection{Building a population model for 2018--2019}
To access Rakiura, visitors need to cross Foveaux Strait using either a scheduled ferry or air service, cruise ship or by a charter aircraft, or other vessel. By drawing on the data sources shown in Table 1 below, which represent the majority (in excess of 90\%) of passenger movements to the island, a population model for the Island was created. This is for the period from 1 April 2018 until 31 March 2019. 

\begin{table}[htp]

\begin{center}
\begin{tabular}{c c c c}
\arrayrulecolor{black}
\hline 
Data Source & Source & Temporal Resolution & Coverage\\
\arrayrulecolor{lightgray}
\hline 
\arrayrulecolor{black}
Rakiura Visitor Levy & Southland District Council & Monthly & Visitors arrival only\\
Ferry Manifest & Real Journeys & Daily & Resident and Visitors\\
Cruise Ship Movements & Maritime NZ & Hourly & Vessels only\\
Population Forecasts & Statistics NZ & - & Residents\\
\hline
\end{tabular}
\end{center}
\label{first_table}
\caption{Population inputs to Rakiura Population Model.}
\end{table}%

In order to build a population model for Rakiura, the net remaining population has been calculated from the Ferry manifested passengers, and assumed 40\% of cruise ship capacity. The later equates to approximately 3,000 cruise ship passengers visiting the Island, which is consistent with historical trends \cite{StatCruise}. To allow for those arriving by scheduled aircraft or chartered service (where no data is held), the net has been pro-rated to match the actual, monthly visitor levy data. 

\subsubsection{Longitudinal temporal analysis for 2016--2019}
A levy from every visitor to Rakiura of \$5 is collected by the aircraft or vessel operator on behalf of the local district council (Southland District Council, SDC). This levy is collected for every domestic and international visitor, with the exception of local ratepayers (including holiday home owners), residents, or visitors who stay for longer than 21 days. SDC has provided this for the period of 2016--2019 on a monthly (aggregated) resolution. 

A number of the VGI and novel indicators reviewed are also aggregated on a monthly basis. Here, we have compared these to the longitudinal data that has been drawn from the Rakiura Visitor Levy, as a proxy for island population. Given the relative stability of the island resident population, this is assumed to be a suitable proxy for changes in island population.  

\subsection{Novel indicators and Volunteered Geographic Information}
The concept of using VGI or novel indicators to understand population fluctuation has been considered by many authors (\citet{dobson2000landscan, nissen2011did, zealand2011evaluation}), however few have had the opportunity to assess in real terms the accuracy or the inherent bias of these methods. Through this study we investigate a range of indicators to assess their suitability to represent wider fluctuations in populations.

Table 2 below shows the indicators we assessed. Two indicators (wastewater and Instagram posts) were assessed on a more fine-grained temporal resolution (daily). Where aggregated datasets were reviewed (i.e., walking track counters and TripAdvisor), they were assessed against the Visitor Levy to understand longer term stability in trends. The assessment of all of these indicators has been undertaken using the raw datasets and no data enrichment has been undertaken.

\begin{table}[htp]

\begin{center}
\begin{tabular}{c c c c c}
\arrayrulecolor{black}
\hline 
Indicator & Aggregated & Type & n\\
\arrayrulecolor{lightgray}
\hline 
\arrayrulecolor{black}
Wastewater volume at Treatment Plant & Daily & Novel Indicators & 365 days\\
Instagram \#StewartIsland uploads & Daily & VGI & 4,027 posts\\
Walking Track Counter & Monthly & Novel Indicators & 36,590 counts\\
TripAdvisor Reviews & Monthly & VGI & 194 reviews\\
\hline
\end{tabular}
\end{center}
\label{first_table}
\caption{Data sources assessed in this review}
\end{table}%

\section{Results}
Presented in Figure 2 below are the key correlations between the daily modelled population for Rakiura and the two daily indicators considered. This clearly shows a strong relationship between daily wastewater volumes and the daily population (Pearson's Correlation of 0.973). This is consistent with existing empirical relationships developed between population and wastewater produced. The strong relationship is likely a function of the lack of significant industry (and industrial effluent) on Rakiura, and the relative simplicity of the network with limited opportunities for water ingress into the network.

The relationship between the population model and \#StewartIsland tagged posts on Social Media platform Instagram is lower (Figure 2). There are a number of complicating factors in this dataset which include: the ability for a user to post multiple posts with the same `tags', the inability to distinguish individual users to identify `individual visits', likely demographic bias (particular in nations where the platform is unavailable), and the volunteered nature of the geographic data (e.g., some users may confuse Rakiura with the South Island of New Zealand). Noting this, there appears to be moderate correlation (0.719) between the modelled population and Instagram posts, and as such opportunities for data enrichment ought to be considered for this dataset. 

\begin{figure}[htbp] \begin{center} 
\resizebox{1.0\textwidth}{!}{ 
	\includegraphics{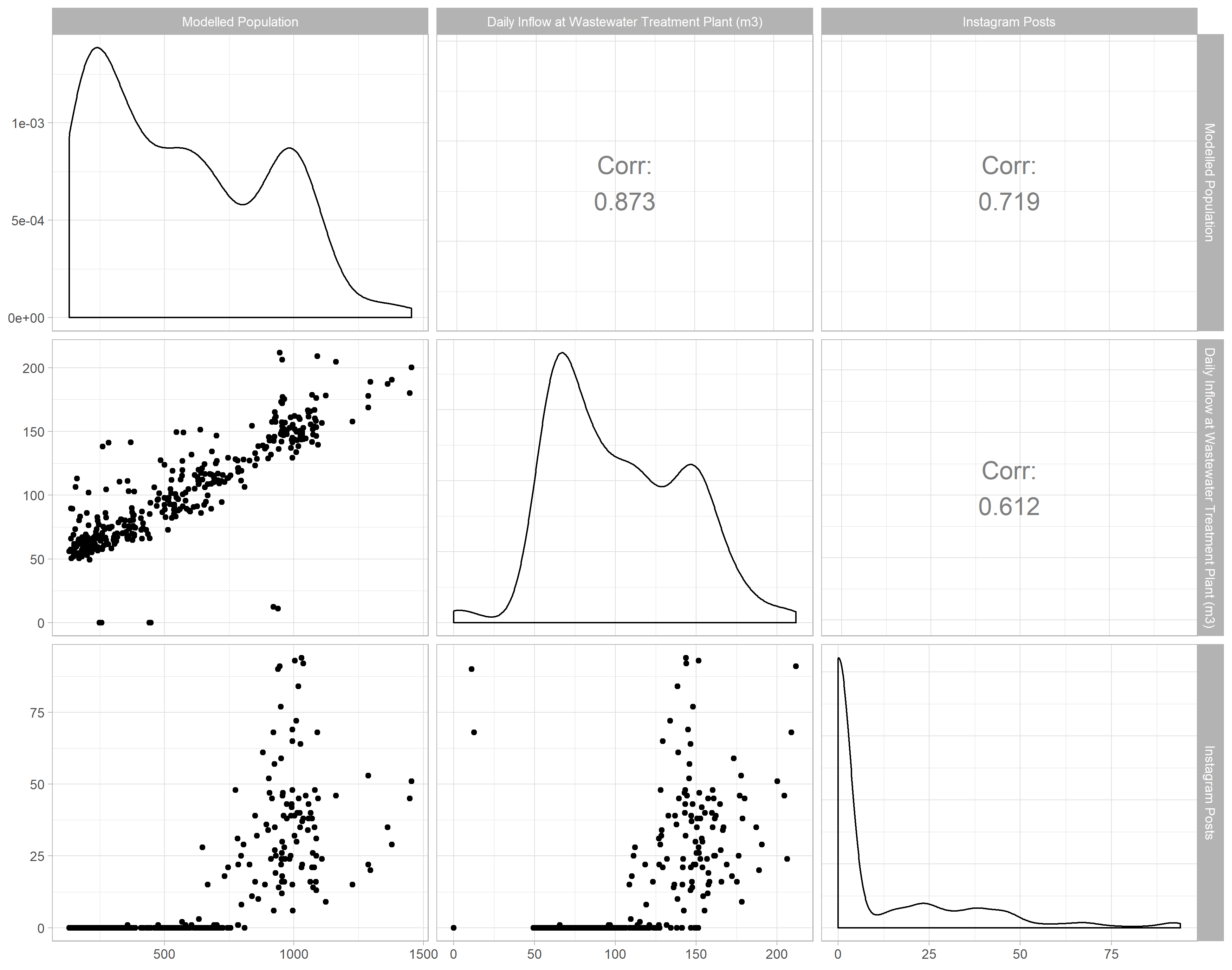}
} \caption{Daily data showing correlations between modelled population, wastewater and Instagram posts between 1 April 2018--31 March 2019} \label{first_figure} \end{center} \end{figure} 
Figure 3 adopts the Southland District Council's visitor levy, as a proxy for population movement. Here on a longer temporal scale (2016--2019) we compare the TripAdvisor reviews for the local `South Seas Hotel' pub and the Department of Conservation (DOC) visitor counter on a local 1hr walking track. In the case of the latter a strong correlation (0.951) is observed between the proxy for the population on Rakiura, and the number of counts on the walking track. This correlation supports the theory, that a large portion of visitors are engaging in nature-based tourism activities.  

TripAdvisor reviews also have a moderate correlation (0.763) to the proxy for population movement. Here it is expected that this will be weighted towards visitors as opposed to holiday home owners or residents. In addition, a visitor is only likely to post once about a particular hotel, activity or restaurant. Further multivariate analysis into understanding particular demographic bias is needed to understand the value of this VGI in population modelling. 

\begin{figure}[htbp] \begin{center} 
\resizebox{1\textwidth}{!}{ 
	\includegraphics{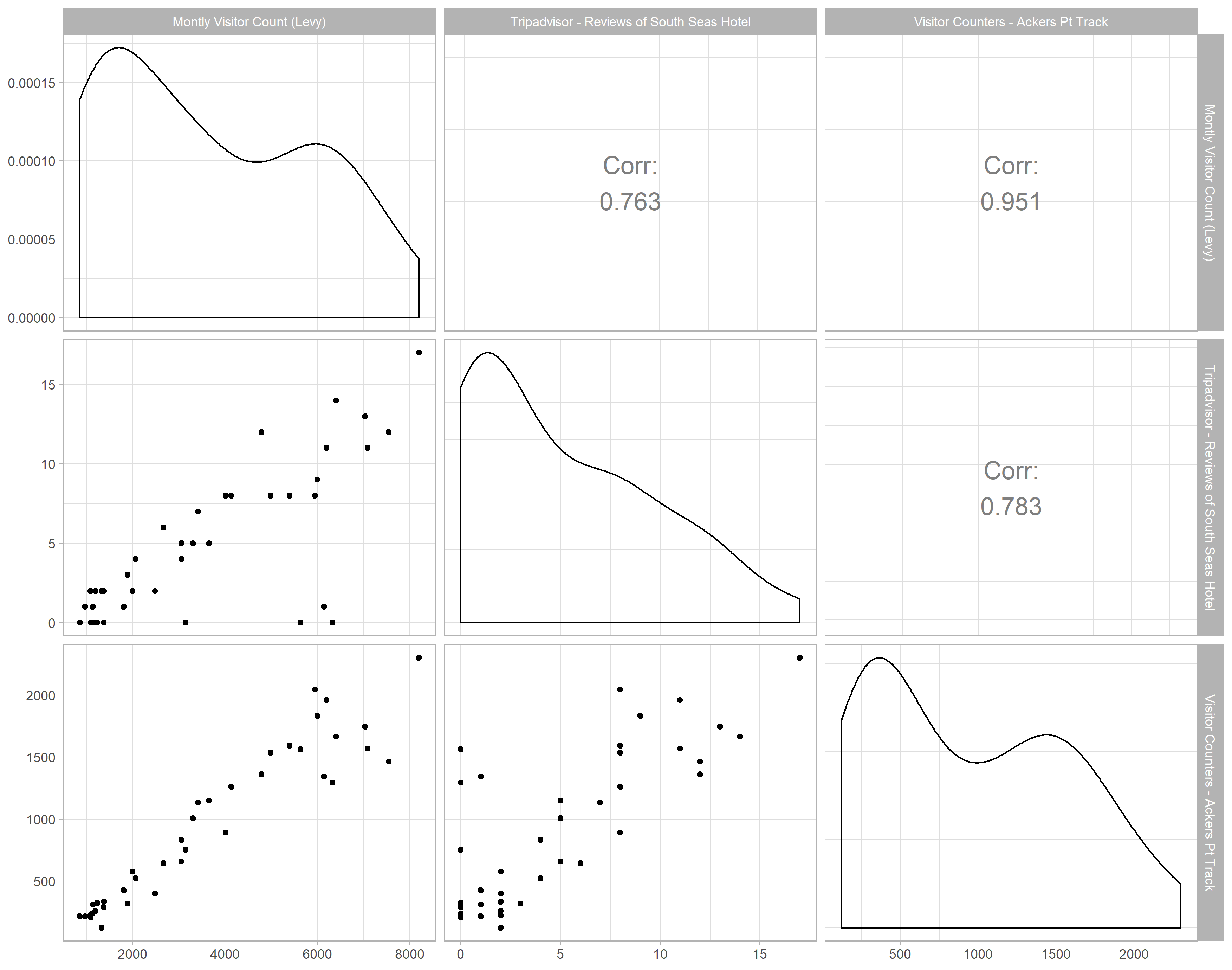}
} \caption{Aggregated monthly data sets showing correlation between measured visitors (Visitor Levy), TripAdvisor reviews of the local restaurant and a visitor counter on a local walking track (between January 2016 and March 2019)} \label{second_figure} \end{center} \end{figure} 

\section{Next steps in applying volunteered geographic information to predicting populations movements}
The research aim of this project was to understand how novel flux indicators can better predict population movement and thus exposure of international and domestic tourists to disaster risk. Research undertaken to date shows that the novel indicators, including VGI have the potential to improve population exposure data informing such risk models.

Moving forward, an assessment of whether data enrichment meaningfully improves the predictive powers of VGI social media data sources is needed. Noting the findings of \citet{middleton2013real} and the limited metadata available in VGI social media data sources, it is proposed that computer vision will be used to ascertain actual geolocation of publicly shared imagery. Preliminary models built in Google's TensorFlow are being tested. Options for data enrichment, including understanding effects of seasonality are being considered. 

Finally, the goal in using the case study of Rakiura, Stewart Island was because it provides an opportunity to compare a number of different kinds of data against ``ground truth'' population numbers. Extending the study to other locations that elicit different kinds of tourism behavior will help establish the generalizability of the findings.

\section{Acknowledgements}
The authors would like to recognise the support of Land Information New Zealand, Resilience to Nature's Challenges - National Science Challenge, QuakeCoRE, and the New Zealand Police Credit Union for their support in this research. We would also like to thank the data providers who have provided data to undertake this research: Wayfare Group Limited, Southland District Council, Maritime New Zealand and the Department of Conservation.

\section{References}

\bibliography{references}

\end{document}